\documentstyle[sprocl]{article}

\input{psfig}

\def\al{\alpha}
\def\be{\beta}

\def\de{\delta}

\def\ve{\varepsilon}
\def\ze{\zeta}

\def\ka{\kappa}
\def\la{\lambda}

\def\rh{\rho}

\def\si{\sigma}

\def\ph{\phi}

\def\ch{\chi}
\def\ps{\psi}

\def\De{\Delta}

\def\Si{\Sigma}

\def\mn{{\mu\nu}}

\def\cl{{\cal L}}

\def\fr#1#2{{{#1} \over {#2}}}
\def\frac#1#2{\textstyle{{{#1} \over {#2}}}}
\def\pt#1{\phantom{#1}}
\def\prt{\partial}

\def\half{{\textstyle{1\over 2}}}
\def\lsim{\mathrel{\rlap{\lower4pt\hbox{\hskip1pt$\sim$}}
    \raise1pt\hbox{$<$}}}
\def\gsim{\mathrel{\rlap{\lower4pt\hbox{\hskip1pt$\sim$}}
    \raise1pt\hbox{$>$}}}

\def\kf{k_{F}}
\def\kfi{(k_{F})_{\ka\la\mu\nu}}
\def\kt{\tilde k}
\def\pht{\hat p}

\def\etal {{\it et al.}}

\newcommand{\beq}{\begin{equation}}
\newcommand{\eeq}{\end{equation}}
\newcommand{\bea}{\begin{eqnarray}}
\newcommand{\eea}{\end{eqnarray}}
\newcommand{\rf}[1]{(\ref{#1})}

\begin{document}

\title{CONSTRAINING LORENTZ VIOLATION USING 
SPECTROPOLARIMETRY OF COSMOLOGICAL SOURCES}

\author{MATTHEW MEWES}

\address{Physics Department, Indiana University,
         Bloomington, IN 47405, U.S.A.\\E-mail: mmewes@indiana.edu} 


\maketitle\abstracts{
Spectropolarimetry of distant sources of electromagnetic
radiation at wavelengths ranging from infrared to 
ultraviolet are used to constrain Lorentz violation.
A bound of $3 \times 10^{-32}$ is placed on
coefficients for Lorentz violation.
}


Lorentz symmetry is an important part
of our current understanding of particle physics.
A violation of this symmetry would be a 
signal of physics beyond the standard model.\cite{cpt98}
For example, Lorentz violation can arise 
in string theory.\cite{kps}
A general Lorentz-violating standard-model
extension has been con\-structed.\cite{ck}
It allows for the possibility that
the remnants of Lorentz violation 
occurring at the  Planck scale may
lead to small violations at energies
attainable today.
A number of experiments have been
performed to test the fermion sector
of the theory.
The CPT-odd coefficients of the photon sector,
which have been constrained
experimentally to a high degree of precision,
are expected to be zero from theoretical considerations.\cite{ck,cptodd}
However, the CPT-even coefficients of the
photon sector have received little attention.
It is the goal of this work to understand the
effects of these coefficients and to place bounds
on them using existing spectropolarimetric
measurements of distant cosmological sources.

A Lorentz-violating electrodynamics can be
extracted from the standard-model extension.
We neglect the CPT-odd coefficients for
the reasons mentioned above.
The relevant Lagrangian is
$\cl = -\frac 1 4 F_{\mu\nu}F^{\mu\nu}
- \frac 1 4 (\kf)_{\ka\la\mu\nu}F^{\ka\la}F^{\mu\nu}$
where $F_\mn$ is the usual field strength,
$F_\mn \equiv \prt_\mu A_\nu -\prt_\nu A_\mu$.
The second term is CPT even and Lorentz violating.
The coefficient $\kfi$ is dimensionless. 
It has the symmetries of the Riemann tensor
and a zero double trace.
This leaves 19 independent components.

The equations of motion resulting from this Lagrangian
are modified inhomogeneous Maxwell equations.
With the aid of the usual homogeneous Maxwell equations,
plane-wave solutions of the form
$F_\mn (x) =F_\mn (p) e^{-ip_\al x^\al}$
can be found.\cite{ck,km}
The modified dispersion relation to leading order 
in $\kfi$ is
\beq
p^0_\pm=(1+\rh\pm\si)\left|\vec p\right|,
\label{kfdispersion}
\eeq
where $\rh=-\half {\kt}_\al^{\pt{\al}\al}$
and $\si^2=\half(\kt_{\al\be})^2-\rh^2$,
with $\kt^{\al\be}\equiv (\kf)^{\al\mu\be\nu}{\pht}_\mu {\pht}_\nu$
and ${\pht}^\mu \equiv {p^\mu}/{|\vec p|}$.
At leading order in $\kfi$, the corresponding
solutions for the electric field, $\vec E_\pm$, are
orthogonal and each is perpendicular to its group velocity 
$\vec v_{g\pm}\equiv \vec\nabla_{\vec p}\ p^0_\pm$.
This implies the unit vectors
$\hat\ve_\pm\equiv \vec E_\pm/|\vec E_\pm|$
form a basis for the electric field.
The general solution is of the form
$\vec E(x) = (E_+ \hat\ve_+e^{-ip^0_+t}+ E_- \hat\ve_-e^{-ip^0_-t}) e^{i\vec p \cdot\vec x}$.
The fact that the phase velocities
$\vec v_{p\pm}\equiv p^0_\pm \vec p /{\vec p}^{\, 2}$
of the two modes differ implies
as the light propagates the relative phase
between modes changes.
The change in relative phase is 
\beq
\De\ph =( p^0_+-p^0_-)t 
\approx 2\pi \De v_p L/\la \approx 4\pi\si L/\la,
\label{deph}
\eeq
where $L$ is the distance the radiation traveled and
$\la$ is its wavelength.
This phase change results in a change in the
polarization of the radiation.
The $L/\la$ dependence suggests, for very distant 
sources producing light at short wavelengths, tiny
differences in phase velocity may become detectable.

It is this $L/\la$ dependence that is exploited
in this work in order to obtain a bound on $\kfi$.
Recent spectropolarimetry of distant galaxies at wavelengths
ranging from infrared to ultraviolet \cite{hough}$^-\,$\cite{vernet}
has made it possible to achieve values of $L/\la$
greater than $10^{31}$.
Measured polarization parameters are typically order 1.
Therefore ,we expect an experimental sensitivity
of $10^{-31}$ or better to components of $\kfi$.

The first step in our analysis is to 
choose a coordinate system in which to work.
A natural choose is a celestial equatorial system
with the 3-axis aligned along the celestial north pole
at equinox 2000.0 at a declination $90^\circ$.
The 1- and 2-axis are at declination $0^\circ$ and right ascension
$0^\circ$ and $90^\circ$, respectively.
The goal is to place bounds on components of $\kfi$ in
this frame.
However, for a source at an arbitrary position on the sky,
this is not the most convenient coordinate system.
Polarization is given by the behavior of $\vec E$ in the
plane perpendicular to the direction of propagation.
Therefore, for each source, we define a 'primed' frame where
$\pht\, '^{\mu}=(1;0,0,1)$ at leading order.
Then the primed-frame basis vector $\hat e_3'$ points from
the source towards the Earth.
To match standard polarimetric conventions we choose
$\hat e_1'$ so that it points south.
The idea is to do much of the analysis in the primed frame
where things are simple and then use observer covariance
to write $(k_F)'_{\ka\la\mu\nu}$ in terms of $\kfi$.
The two frames are related by a rotation.

In the primed frame 
$\rh=\half(\kt '^{\, 11}+\kt '^{\, 22})$
and
$\si^2=(\kt '^{\, 12})^2
+\fr 1 4 (\kt '^{\, 11}-\kt '^{\, 22})^2$.
The form of $\si^2$ suggests defining an angle $\xi$ such that 
$\kt '^{\, 12} = \si \sin\xi$ and 
$\half(\kt '^{\, 11}-\kt '^{\, 22})= \si \cos\xi$.
Note that while $\rh$ and $\si$ are frame independent,
$\xi$ is not.
Solving the modified Maxwell equations in this frame
gives $\hat\ve_\pm\propto(\sin\xi,\pm1-\cos\xi,0)$.
This implies that the birefringent modes are linearly
polarized.
From the solutions $\hat\ve_\pm$ and Eq.\ \rf{deph} it is evident
that $\si$ and $\xi$ are the relevant parameters for
polarimetry of a particular source.
More precisely, $\si\sin\xi$ and $\si\cos\xi$ represent the minimal
linear combinations of $\kfi$ effecting the polarization
of a given source.
It can be shown that $\si\sin\xi$ and $\si\cos\xi$, written in terms
of the celestial equatorial $\kfi$, depend on the right ascension and
declination of the source and 10 independent components of $\kfi$.
It is these ten components that are bounded in this work.
We denote these components as $k^a, a=1,...,10$.
A suitable choose for $k^a$ in terms of $\kfi$
can be found in the literature.\cite{km}

In general, plane waves are elliptically polarized.
The ellipse can be characterized by two angles:
$\psi$, the angle between $\hat e'_1$ and the major axis of the ellipse
and $\ch=\pm\arctan\frac{minor\ axis}{major\ axis}$, which describes
the shape of the ellipse and the helicity.
The change in relative phase, Eq.\ \rf{deph}, results in a change
in both these angles.
Most published polarimetric data of astronomical sources
do not include measurements of $\ch$.
Therefore, our analysis focuses on finding an expression for the
change in $\psi$.
It will not only depend on $k^a$, the wavelength $\la$,
and the distance to the source $L$, but also on the values of 
$\psi$ and $\ch$ when the light is emitted.
Our approach is to look for wavelength dependence in the observed
polarization.
This approach assumes that the polarization at the source
is relatively constant over the wavelengths considered.

We seek an expression for $\de\psi=\psi-\psi_0$,
the difference between $\psi$ at two wavelengths, $\la$ and $\la_0$.
We find 
\beq
\de\psi
=\half\tan^{-1}{\fr
{\sin\tilde\xi\cos\ze_0+\cos\tilde\xi\sin\ze_0\cos(\de\ph-\ph_0)}
{\cos\tilde\xi\cos\ze_0-\sin\tilde\xi\sin\ze_0\cos(\de\ph-\ph_0)}},
\label{dpsi}
\eeq
where $\de\ph=4\pi\si(L/\la-L/\la_0)$,
$\tilde\xi=\xi-2\psi_0$ and 
$\ph_0\equiv \tan^{-1}(\tan2\ch_0/\sin\tilde\xi)$,
$\ze_0\equiv \cos^{-1}(\cos2\ch_0\cos\tilde\xi)$.\cite{km}
The polarization at $\la_0$ is given by $\ps_0$ and $\ch_0$.
Two of these parameters need to be fit to the data.
This is equivalent to fitting the initial polarization.
The third parameter can be fixed to a convenient value.

Table 1 lists 16 sources with published polarimetric data
with observed wavelengths ranging from 400 to 2200 nm.
For each source, we choose $\psi_0$ as the mean
polarization angle and use Eq.\ \rf{dpsi} to
create a $\chi^2$ distribution.
Each distribution is a function of $\psi$, $\tilde\xi$,
$\la_0$, and $\chi_0$.
They are then minimized with respect to $\la_0$, and $\chi_0$.

Figure 1 shows the minimized distribution for the 
source 3CR 68.1.
The features of this contour are common to all
sources in Table 1.
The contour corresponds to a confidence level of
about $(100-10^{-9})\%$.
We see from Fig. 1 that the parameter space away
from $\si=0$, $\tilde\xi=0^\circ, \pm90^\circ$ are
eliminated by this source.
These are only the regions where the theory
predicts no change in $\psi$.
The regions near $\tilde\xi=0^\circ, \pm90^\circ$
correspond to the radiation being in a specific
combination of birefringent modes.
For example, $\tilde\xi=0^\circ$ occurs if
the light is emitted in only one mode.
We assume the probability of this happening for
all 16 sources is small.
With this assumption, the $\chi^2$ can be used
to place a conservative constraint on $\si$.
In Fig. 1, the bound is shown as a horizontal line.

\begin{center}
\begin{tabular}{|l||c|c|c|}
\hline
\multicolumn{1}{|c||}{Source} 
& $L$~(Gpc) & $10^{30}L/\la$ & $\log_{10}\si$ \\ 
\hline \hline
IC 5063 \cite{hough}                      & 0.04 & 0.56 - 2.8 & -30.8 \\
3A 0557-383 \cite{brindle}                & 0.12 & 2.2 - 8.4  & -31.2 \\
IRAS 18325-5925 \cite{brindle}            & 0.07 & 1.0 - 4.9  & -31.0 \\
IRAS 19580-1818 \cite{brindle}            & 0.13 & 1.8 -  9.1 & -31.0 \\
3C 324 \cite{cimatti465}                  & 1.69 & 58 - 130   & -32.2 \\
3C 256 \cite{dey}                         & 1.92 & 70 -  140  & -32.2 \\
3C 356 \cite{cimatti476}                  & 1.62 & 57 - 120   & -32.2 \\
F J084044.5+363328 \cite{brothertonfirst} & 1.71 & 62 - 120   & -32.2 \\
F J155633.8+351758 \cite{brothertonfirst} & 1.82 & 67 - 110   & -32.2 \\
3CR 68.1 \cite{brotherton}                & 1.70 & 59 - 130   & -32.2 \\
QSO J2359-1241 \cite{brothertonqso}       & 1.48 & 87 - 90    & -31.1 \\
3C 234 \cite{kishimoto}                   & 0.55 & 51 - 75    & -31.7 \\
4C 40.36 \cite{vernet}                    & 2.02 & 73 - 160   & -32.2 \\
4C 48.48 \cite{vernet}                    & 2.04 & 75 - 160   & -32.2 \\
IAU 0211-122 \cite{vernet}                & 2.04 & 74 - 160   & -32.2 \\
IAU 0828+193 \cite{vernet}                & 2.08 & 78 - 160   & -32.2 \\
\hline
\end{tabular}
\end{center}
\begin{center}
Table 1. Source Data. 
\end{center}

\begin{figure}[hb]
\centerline{
\psfig{figure=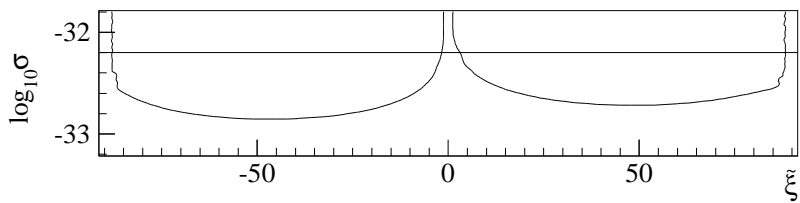,width=0.9\hsize}}
\caption{
Contours of $\ch^2$ for the source 3CR 68.1.}
\label{fig1}
\end{figure}

The bounds for each source are listed in the last
column of Table 1.
To estimate the constraint on $k^a$, we assume,
for each source, the data are consistent with $\si=0$.
The bounds can be thought of as an estimate of the
error $\de\ph$ in a null measurement.
We construct a second $\chi^2$ distribution,
$\chi^2=\Si_j(\si_j)^2/(\de\si_j)$,
where the sum ranges over the 16 sources.
This $\chi^2$ is a quadratic form in the $k^a$
coefficients.
A constant value of $\chi^2$
correspond to a ten-dimensional ellipsoid in the
$k^a$ space.
We place a bound on the magnitude $|k^a|=\sqrt{k^ak^a}$
by minimizing $\chi^2$ with respect to the other nine
degrees of freedom.
This yields a bound of $|k^a|<3\times10^{-32}$ at the
90\% confidence level.

These are the first bounds on the coefficients $\kfi$.
They are comparable to the best existing bounds
in the fermion sector of the standard-model extension.
An improvement in this bound can be expected if more
measurements similar to those used here are made.
Similar measurements of $\chi$ could also be used to
improve the bound.
In the future it may be possible to include X-ray
polarimetry,\cite{costa}
which may lead to an improvement of
several orders of magnitude.


\end{document}